\documentclass[12pt]{article}

\begin{document}

\baselineskip 0.85cm
\topmargin -0.55in
\oddsidemargin -0.1in

\let\ni=\noindent

\renewcommand{\thefootnote}{\fnsymbol{footnote}}

\newcommand{\CKM}{Cabibbo--Kobayashi--Maskawa}

\newcommand{\SM}{Standard Model }

\pagestyle {plain}

\setcounter{page}{1}

\pagestyle{empty}

\vspace{0.3cm}

\begin{flushright}
{IFT-00/15\\
hep-ph/0007255}
\end{flushright}
\vspace*{5mm}

{\large\centerline{\bf Two--mixing texture for three active neutrinos{\footnote {Supported in part by the Polish KBN--Grant 2 P03B 052 16 (1999--2000).}}}}

\vspace{0.8cm}

{\centerline {\sc Wojciech Kr\'{o}likowski}}

\vspace{0.8cm}

{\centerline {\it Institute of Theoretical Physics, Warsaw University}}

{\centerline {\it Ho\.{z}a 69,~~PL--00--681 Warszawa, ~Poland}}

\vspace{0.5cm}

{\centerline{\bf Abstract}}

\vspace{0.3cm}

The conjecture that among three massive neutrinos $ \nu_1\,, \,\nu_2\,, \,\nu_3$ there is no direct mixing between $ \nu_1$ and $ \nu_3 $ leads to a two--mixing texture for three active neutrinos $ \nu_e\,, \,\nu_\mu\,, \,\nu_\tau$. This texture, much discussed previously, is neatly consistent with the observed deficits of solar $ \nu_e$'s  and atmospheric $\nu_\mu $'s, but (without extra mixing with at least one sterile neutrino $ \nu_s $) predicts no LSND effect for accelerator $ \nu_\mu$'s. In this option, the masses $ m^2_1\stackrel{<}{\sim} m^2_2 \ll m^2_3$ are readily estimated. The characteristic feature of the two--mixing neutrino texture that {\it only the close neighbours in the hierarchy of massive neutrinos $ \nu_1\,, \,\nu_2\,, \,\nu_3$ mix significantly} may be physically meaningful. Going out from the notion of mixing matrix we construct an intrinsic occupation--number operator whose {\it eigenvalues} 0, 1, 2 {\it numerate the three generations of massive neutrinos}. Analogical constructions work also for charged leptons as well as for up and down quarks.

\vspace{0.2cm}

\ni PACS numbers: 12.15.Ff , 14.60.Pq , 12.15.Hh .

\vspace{0.5cm}

\ni July 2000

\vfill\eject

~~~
\pagestyle {plain}

\setcounter{page}{1}

If one conjectures that in the generic \CKM$\!$--type matrix for leptons [1],

\begin{equation} 
U =  \left( \begin{array}{ccl} c_{13}c_{12}  & c_{13} s_{12}  & s_{13} e^{-i \delta}  \\ -c_{23} s_{12} - s_{13} s_{23} c_{12}  e^{i \delta} &  \;c_{23} c_{12} - s_{13} s_{23} s_{12}  e^{i \delta}   & c_{13} s_{23}   \\ \;s_{23} s_{12} - s_{13} c_{23} c_{12}  e^{i \delta}  & -s_{23} c_{12} - s_{13} c_{23} s_{12}  e^{i \delta}  & c_{13} c_{23} \end{array} \right) 
\end{equation} 

\ni with $s_{ij} = \sin \theta_{ij} > 0 $ and $c_{ij} = \cos \theta_{ij} \geq 0 $, $ (i\,,\,j = 1,2,3)$, there is practically no direct mixing of massive neutrinos $\nu_1 $ and $\nu_3 $ ({\it i.e.}, $\theta_{13} = 0$), then $ U $ is reduced to the following two--mixing form much discussed previously [2]:

\begin{equation} 
U =  \left( \begin{array}{ccc} c_{12} & s_{12}  & 0  \\ -c_{23} s_{12}  &  \;c_{23} c_{12} & s_{23} \\ \;s_{23} s_{12} & -s_{23} c_{12} & c_{23} \end{array} \right) =  \left( \begin{array}{ccc} 1 & 0  & 0  \\ 0 & \; c_{23} & s_{23} \\ 0 & -s_{23} & c_{23} \end{array} \right)  \left( \begin{array}{ccc} \;c_{12} & s_{12}  & 0  \\ -s_{12} & c_{12} & 0 \\ 0 & 0 & 1 \end{array} \right) \;. 
\end{equation} 

\ni For the two--mixing option (2) the neutrino mixing formula $\nu_\alpha = \sum_i U_{\alpha i} \nu_i $ takes the form

\begin{eqnarray}
\nu_{e} & = & c_{12} \nu_1 + s_{12} \nu_2 \;, \nonumber \\
\nu_{\mu} & = & c_{23}(-s_{12}\nu_1 + c_{12} \nu_2) + s_{23} \nu_3 \;, \nonumber \\
\nu_{\tau} & = & -s_{23}(-s_{12}\nu_1 + c_{12} \nu_2) + c_{23} \nu_3 \;, 
\end{eqnarray}

\ni while the inverse neutrino mixing formula $\nu_i = \sum_\alpha U^*_{\alpha i} \nu_\alpha $ gives 

\begin{eqnarray}
\nu_{1} & = & c_{12} \nu_e - s_{12}(c_{23} \nu_\mu - s_{23}\nu_\tau) \;, \nonumber \\
\nu_{2} & = & s_{12}\nu_e + c_{12}(c_{23} \nu_\mu - s_{23} \nu_\tau) \;, \nonumber \\
\nu_{3} & = & s_{23}\nu_\mu + c_{23} \nu_\tau \;. 
\end{eqnarray}

In the representation, where the charged--lepton mass matrix is diagonal (and thus the corresponding diagonalizing matrix --- unit), the lepton mixing matrix $ U = \left( U_{\alpha i} \right) $ $(\alpha = e \,,\, \mu \,,\, \tau\;,\; i = 1,2,3)$ is, at the same time, the diagonalizing matrix for neutrino mass matrix $ M = \left( M_{\alpha \beta} \right) $ $(\alpha\,,\,\beta = e \,,\, \mu \,,\, \tau)$ , $ U^\dagger M U = {\rm diag}(m_1\,,\,m_2\,,\,m_3)$ with  $ m^2_1 \leq m^2_2 \leq m^2_3 $, so that $ M = \left( \sum_i U_{\alpha i} U^*_{\beta i} m_i \right)$. In this case, the orthogonal two--mixing form (2) of $ U $ leads to the real and symmetric

\begin{equation} 
\!M\!\! = \!\!  \left(\! \begin{array}{ccc} c_{12}^2 m_1\! +\! s^2_{12} m_2  & (m_2 \!-\! m_1) c_{12} s_{12}c_{23}  & -(m_2 \!-\! m_1) c_{12} s_{12} s_{23} \\ (m_2 \!-\! m_1) c_{12} s_{12}c_{23} &  s^2_{23}m_3 \!+\! c^2_{23}( s^2_{12}m_1 \!+\! c^2_{12} m_2) & ( m_3 \!-\! s^2_{12}m_1 \!-\! c^2_{12}m_2)c_{23} s_{23} \\ -\! (m_2 \!-\! m_1) c_{12} s_{12}s_{23}  & ( m_3 \!-\! s^2_{12}m_1 \!-\! c^2_{12}m_2) c_{23} s_{23} & c^2_{23} m_3 \!+\! s^2_{23}(s^2_{12} m_1 \!+\! c^2_{12} m_2) \end{array}\!\! \right) .
\end{equation} 

\ni Here, as is seen from Eq. (4), the values $ c_{23} = 1/\sqrt{2} = s_{23}$ give maximal mixing of $ \nu_\mu $ and $ \nu_\tau $: $( \nu_\mu \pm  \nu_\tau )/\sqrt{2}$, and then $ c_{12} \simeq 1/\sqrt{2} \simeq s_{12}$ --- a nearly maximal mixing of $ \nu_e $ and $ (\nu_\mu - \nu_\tau)/\sqrt{2} $: approximately $[\nu_e \pm  (\nu_\mu - \nu_\tau)/\sqrt{2}]/\sqrt{2} $. 

From the familiar neutrino oscillation formulae

\begin{equation} 
P(\nu_\alpha \rightarrow \nu_\beta) = |\langle \nu_\beta| e^{iPL}|
\nu_\alpha\rangle|^2 = \delta_{\alpha \beta} - 4 \sum_{j>i} U^*_{ \beta j} U_{\alpha j}U_{\beta i} U^*_{ \alpha i} \sin^2 x_{j i} \; ,
\end{equation} 

\ni with

\begin{equation} 
x_{j i} = 1.27 \frac{\Delta m^2_{j i} L}{E}\; , \; \Delta m^2_{j i} = m^2_j - m^2_i
\end{equation}

\ni  ($\Delta m^2_{j i}$, $L$ and $E$ measured in eV$^2$, km and GeV, respectively) which is valid for $ U^*_{ \beta j} U_{\alpha j}U_{\beta i} U^*_{ \alpha i}$ real (CP violation neglected), one infers in the case of two--mixing option (2) that

\begin{eqnarray}
P(\nu_e \rightarrow \nu_e) & = & 1 - (2 c_{12} s_{12})^2  \sin^2 x_{21} \;,
\nonumber \\ 
P( \nu_\mu \rightarrow \nu_\mu) & = & 1 - (2 c_{12} s_{12} c_{23})^2  \sin^2 x_{21} - (2 c_{23} s_{23})^2( s^2_{12}  \sin^2 x_{31} +c^2_{12}  \sin^2 x_{32})    \nonumber \\ 
& \simeq & 1 - (2 c_{23} s_{23})^2  \sin^2 x_{32} \; , \nonumber \\ 
P(\nu_\mu \rightarrow \nu_e) & = &  (2 c_{12} s_{12} c_{23})^2  \sin^2 x_{21} \;,
\end{eqnarray}

\ni where the final step in the second formula is valid when $ \Delta m^2_{21} \ll \Delta m^2_{31} \simeq \Delta m^2_{32}$ or equivalently $ m^2_1 \simeq m^2_2 \ll m^2_3 $.

The first formula (8) is consistent with the observed deficit of solar $ \nu_e $'s if one applies the vacuum global solution or large--angle MSW global solution or finally LOW global solution [3] with $(2 c_{12} s_{12})^2 \leftrightarrow \sin^2 2\theta_{\rm sol} \sim (0.90 $ or 0.79 or 0.91) and $ \Delta m^2_{21} \leftrightarrow \Delta m^2_{\rm sol} \sim (4.4 \times 10^{-10} $ or $2.7 \times 10^{-5}$ or $1.0 \times 10^{-7}) \, {\rm eV}^2$, respectively. This gives $ c^2_{12} \sim 0.5 + (0.16 $ or 0.23 or 0.15) and $ s^2_{12} \sim 0.5 - (0.16 $ or 0.23 or 0.15), when taking $ c^2_{12} \geq s^2_{12} $.

The second formula (8) describes correctly the observed deficit of atmospheric $ \nu_\mu $'s [4] if $(2 c_{23} s_{23})^2 \leftrightarrow \sin^2 2 \theta_{\rm atm} \sim 1$ and $ \Delta m^2_{32} \leftrightarrow \Delta m^2_{\rm atm} \sim 3.5 \times 10^{-3} \, {\rm eV}^2$, since then $ \Delta m^2_{21} \ll \Delta m^2_{31} \simeq \Delta m^2_{32}$ for $ \Delta m^2_{21}$ determined as in the case of solar $ \nu_e $'s. This implies that $c^2_{23} \sim 0.5 \sim s^2_{23}$  and $ m^2_3 \sim 3.5 \times 10^{-3}\,{\rm eV}^2 $, because $ m^2_{1} \simeq m^2_{2} \ll m^2_{3}$.

Then, the third formula (8) shows that no LSND effect for accelerator $ \nu_\mu $'s [5] should be observed, $ P(\nu_\mu \rightarrow \nu_e) \sim 0 $, since with $ \Delta m^2_{21} \leftrightarrow \Delta m^2_{\rm sol} \sim (10^{-10 }$ or $ 10^{-5}$ or $ 10^{-7})\, {\rm eV}^2 \ll \Delta m^2_{\rm LSND}\sim 1 \, {\rm eV}^2 $, one gets $ \sin^2(x_{12})_{\rm LSND} \sim 10^{-19}$ or $10^{-9 } $ or $10^{-14} \ll \sin^2 x_{\rm LSND} \sim 1 $, while $(2 c_{12} s_{12} c_{23})^2 \sim (0.90 $ or 0.79 or $ 0.91) \times 0.5 >\sin^2 2\theta_{\rm LSND} \sim 10^{-2}$. As is well known, the confirmation of LSND effect would require (beside three active neutrinos $ \nu_e\,,\,\nu_\mu\,,\,\nu_\tau $) the existence of at least one sterile neutrino $\nu_s $ (blind to all \SM gauge interactions) [6,7], mixing with $\nu_e $ through a mass generation mechanism.

In the case of Chooz experiment looking for oscillations of reactor $\bar{ \nu}_e$'s [8], where it happens that $ (x_{32})_{\rm Chooz} = 1.27 \Delta m^2_{32} L_{\rm Chooz}/E_{\rm Chooz} \sim 1 $ for $\Delta m^2_{32} \leftrightarrow \Delta m^2_{\rm atm} $, the first formula (8) leads to $ P(\bar{\nu}_e \rightarrow \bar{\nu}_e)  \sim 1 $, since $ (x_{21})_{\rm Chooz} \ll (x_{32})_{\rm Chooz} \sim 1$ for $\Delta m^2_{21} \leftrightarrow \Delta m^2_{\rm sol} $ ( $ U_{e3} = 0 $ in our case). This is consistent with the negative result of Chooz experiment. We can see, however, that for the actual lepton counterpart of \CKM matrix the entry $ U_{e3}$ may be a potential correction to the two--mixing option (2) ($ |U_{e3}| < 0.2 $ according to the estimation in Chooz experiment).

Further on, we will put $ c_{23} = 1/\sqrt{2} = s_{23}$. Then, from Eq. (5) we infer that

\begin{equation} 
M \!=\!  \left( \!\! \begin{array}{ccc} c_{12}^2 m_1 \!+\! s^2_{12} m_2  & (m_2 \!-\! m_1) c_{12} s_{12}/ \sqrt{2} & -(m_2 \!-\! m_1) c_{12} s_{12} / \sqrt{2}   \\ (m_2 \!-\! m_1) c_{12} s_{12}/ \sqrt{2} &  (m_3 \!+\!  s^2_{12}m_1 \!+\! c^2_{12} m_2)/2 & ( m_3 \!-\! s^2_{12}m_1 \!-\! c^2_{12}m_2)/2 \\ -(m_2 \!-\! m_1) c_{12} s_{12} / \sqrt{2} & ( m_3 \!-\! s^2_{12}m_1 \!-\! c^2_{12}m_2) /2 & (m_3 \!+\! s^2_{12} m_1 \!+\! c^2_{12} m_2) / 2 \end{array}\!\! \right) \;.
\end{equation} 

\ni Here, $ M_{e \mu} = - M_{e \tau}$, $ M_{\mu \mu} = M_{\tau \tau}$ and 

\begin{eqnarray} 
 M_{e e} = c^2_{12}m_1 + s^2_{12} m_2 \;,\; M_{e e} & \!\! + \!\! & M_{\mu \mu} - M_{\mu \tau} = m_1 + m_2 \;,\; M_{\mu \mu} + M_{\mu \tau} = m_3 \;,\nonumber \\  M_{e \mu} & \!\! = \!\! & (m_2 - m_1)c_{12} s_{12}/\sqrt{2}\;.
\end{eqnarray}

Assuming that $ M_{e e} = 0 $, we get from Eq. (10) the relations $ M_{\mu  \mu} = (m_3 + m_2 + m_1)/2 $, $ M_{\mu  \tau} = (m_3 - m_2 - m_1)/2 $, $ M_{e  \mu} = (s_{12}/c_{12}) m_2/ \sqrt{2} $, and

\begin{equation} 
\frac{m_1}{m_2} = - \frac{s^2_{12}}{c^2_{12}}\; , \; \Delta m^2_{21} \equiv m^2_2 - m^2_1 = m^2_2 \frac{c^2_{12} - s^2_{12}}{c^4_{12}}
\end{equation}

\ni or

\begin{equation} 
m_1 = - \sqrt{\Delta m^2_{21}} \frac{s^2_{12}}{\sqrt{c^2_{12} - s^2_{12}}}\; , \; 
m_2 =  \sqrt{\Delta m^2_{21}} \frac{c^2_{12}}{\sqrt{c^2_{12} - s^2_{12}}}\; , 
\end{equation}

\ni when taking $ m_1 \leq m_2 $. For instance, applying to Eq. (12) the LOW solar solution [3] {\it i.e.}, $ s^2_{12} \sim 0.5 - 0.15 $, $ c^2_{12} \sim 0.5 + 0.15 $ and $ \Delta m^2_{21} \sim 1.0\times 10^{-7}\,{\rm eV}^2 $, we estimate

\begin{equation} 
m_1  \sim - 2.0 \times 10^{-4}\,{\rm eV}\; , \; m_2 \sim 3.8 \times 10^{-4}\,{\rm eV}\;  , 
\end{equation}

\ni while the Super--Kamiokande result $ \Delta m^2_{32} \sim 3.5 \times 10^{-3}\,{\rm eV}^2 $ [4] leads to the estimation

\begin{equation} 
m_3  \sim 5.9 \times 10^{-2}\,{\rm eV}\; , 
\end{equation}

\ni what shows explicitly that $ | m_1| \stackrel{<}{\sim} m_2 \ll m_3 $. Thus,    in this case

\vspace{-0.2cm}

\begin{equation} 
M_{e e} =0 , M_{\mu \mu} = M_{\tau \tau}  \sim 3.0 \times 10^{-2}{\rm eV} , M_{e \mu} =  -M_{e \tau} \sim 1.9 \times 10^{-4}{\rm eV}, M_{\mu \tau} \sim 3.0 \times 10^{-2}{\rm eV},
\end{equation}

\ni where $  M_{\mu \mu} \stackrel{>}{\sim} M_{\mu \tau} \gg M_{e \mu}$.

In the general case, carrying out the diagonalization of mass matrix $ M = \left( M_{\alpha \beta} \right) $ given in Eq. (9), we obtain

\begin{eqnarray} 
 m_{1,2} & = & \frac{M_{e e} +  M_{\mu \mu} -  M_{\mu \tau}}{2} \mp \sqrt{ \left( \frac{ M_{e e} - M_{\mu \mu} + M_{\mu \tau}}{2} \right)^2 + 2 M^2_{e \mu} } \nonumber \\ & = & \left\{ \begin{array}{l} M_{e e} - X  M_{e \mu} \sqrt{2} \\  M_{\mu \mu} -  M_{\mu \tau}+ X  M_{e \mu} \sqrt{2} \end{array} \right. \;, \nonumber \\ 
m_3 & = &  M_{\mu \mu} + M_{\tau \tau}
\end{eqnarray} 

\ni and

\begin{equation} 
c_{12} = \frac{1}{\sqrt{1+ X^2}}\;,\; s_{12} = \frac{X}{\sqrt{1 + X^2}}\; , 
\end{equation}

\ni where

\begin{eqnarray} 
X & \equiv & -\frac{m_1 - M_{e e}}{M_{e \mu} \sqrt{2}} =\frac{M_{e \mu} \sqrt{2}}{m_2 - M_{e e}} \nonumber \\ & = & \frac{ M_{e e} - M_{\mu \mu} + M_{\mu \tau}}{2 M_{e \mu} \sqrt{2}} +  \sqrt{ \left( \frac{ M_{e e} - M_{\mu \mu} + M_{\mu \tau}}{2 M_{e \mu} \sqrt{2}} \right)^2 + 1}\, > 0\;.
\end{eqnarray} 

\ni Here, $ 0 < X < 1 $ if $ M_{e e} - M_{\mu \mu} + M_{\mu \tau} <0 $. For instance, for LOW solar solution [3]

\begin{equation} 
X = \frac{s_{12}}{c_{12}} \sim \sqrt{0.54} = 0.73 \; , 
\end{equation}

\ni showing that then $ M_{e e} - M_{\mu \mu} + M_{\mu \tau} <0 $.

In conclusion, the two--mixing texture of three (Dirac or Majorana) active neutrinos $ \nu_\alpha \;\,(\alpha = e\,,\,\mu\,,\,\tau) $, described by the formulae (2) and (5), is neatly consistent with the observed solar and atmospheric neutrino deficits, but it predicts no LSND effect whose confirmation should imply, therefore, the existence of at least one sterile neutrino $ \nu_s $, mixing with $ \nu_e $. This might be either one extra,  light (Dirac or Majorana) sterile neutrino $ \nu_s $ or one of three conventional, light Majorana sterile neutrinos $ \nu_\alpha^{(s)} = \nu_{\alpha R} + (\nu_{\alpha R})^c\;\;(\alpha = e\,,\,\mu\,,\,\tau)$  existing in this case beside three light Majorana active neutrinos $\nu_\alpha^{(a)} = \nu_{\alpha L} + (\nu_{\alpha L})^c \;\;(\alpha = e\,,\,\mu\,,\,\tau)$ [of course, $\nu_\alpha^{(a)} = \nu_{\alpha L}$ and $ \nu_{\alpha L}^{(s)} = (\nu_{\alpha R})^c] $.

The essential agreement of the observed neutrino oscillations with the two--mixing option (2) for $ U $ (provided there is really no LSND effect) suggests that the conjecture of absence of direct mixing of massive neutrinos $ \nu_1 $ and $ \nu_3 $, leading to $ U $ of the form (2), is somehow physically important. This absence tells us that {\it only the close neighbours in the hierarchy of massive neutrinos $ \nu_1\,, \,\nu_2\,, \,\nu_3 $ mix significantly}.

Making use of Gell--Mann matrices (in the space of three generations)

\begin{equation} 
\lambda_2 =\left(  \begin{array}{ccc} 0 & -i & 0 \\ i & 0 & 0  \\ 0 & 0 & 0 \end{array} \right) \;,\; 
\lambda_7 = \left( \begin{array}{ccc} 0 & 0 & 0 \\ 0 & 0 & -i  \\ 0 & i & 0 \end{array} \right)
\end{equation} 

\ni we can rewrite the two--mixing matrix (2) in the compact form

\begin{equation} 
U = U^{(23)} U^{(12)} = e^{i \lambda_7 \theta_{23}}\,e^{i \lambda_2 \theta_{12}}\;,
\end{equation} 

\ni while the generic matrix (1) includes also the phased 13--rotation

\begin{equation} 
\!\!U^{(13)} \!= \! \left( \! \begin{array}{ccc} c_{13} & 0 & s_{13}e^{-i\delta} \\ 0 & 1 & 0  \\ -s_{13}e^{i\delta}  & 0 & c_{13} \end{array} \right)\! = \!\left( \! \begin{array}{ccc} 1 & 0 & 0 \\ 0 & 1 & 0  \\ 0 & 0 & e^{i \delta} \end{array} \right) e^{i \lambda_5 \theta_{13}} \left( \!\begin{array}{ccc} 1 & 0 & 0 \\ 0 & 1 & 0  \\ 0 & 0 & e^{-i\delta} \end{array} \!\right)  , 
\lambda_5 \! = \!\left( \!\begin{array}{ccc} 0 & 0 & -i  \\ 0 & 0 & 0 \\  i & 0 & 0 \end{array}\! \right) ,
\end{equation} 

\ni inserted between the previous 23-- and 12--rotations, $ U^{(23)}$ and $ U^{(12)}$, of closely neighbouring massive neutrinos, $ U = U^{(23)} U^{(13)} U^{(12)} $ .

Then, in terms of the (truncated) annihilation and creation operators (in generation space)

\begin{equation} 
a = \left(  \begin{array}{ccc} 0 & 1 & 0 \\ 0 & 0 & \sqrt{2}  \\ 0 & 0 & 0 \end{array} \right) \;,\; 
a^\dagger = \left( \begin{array}{ccc} 0 & 0 & 0 \\ 1 & 0 & 0  \\ 0 & \sqrt{2} & 0 \end{array} \right)
\end{equation} 

\ni satisfying together with the operator

\begin{equation} 
n = a^\dagger a =\left(  \begin{array}{ccc}  0 & 0 & 0 \\ 0 & 1 & 0 \\ 0 & 0 & 2 \end{array} \right)
\end{equation} 

\ni the canonical annihilation and creation relations

\begin{equation} 
[a\,,\,n] = a \;\; , \;\; [a^\dagger\,,\,n] = - a^\dagger \; ,
\end{equation} 

\ni but obeying also the truncation condition

\begin{equation} 
a^3 = 0 = a^{\dagger\,3} \; ,
\end{equation} 

\ni we can put

\begin{equation} 
\lambda_2 = \frac{1}{2i} a(a - a^\dagger)a^\dagger\; ,\; \lambda_7 = \frac{1}{i \sqrt{2}} a^\dagger(a - a^\dagger)a \; ,\; \lambda_5 = \frac{1}{i \sqrt{2}} (a^2 - a^{\dagger\,2}) 
\end{equation} 

\ni in exponents of the factor matrices $ U^{(12)}\,,\,U^{(23)}\,,\,U^{(13)} $, respectively. Other Gell--Mann matrices, absent from $ U $, can be put in the form

\begin{equation} 
\lambda_1 = \frac{1}{2} a(a + a^\dagger)a^\dagger\; ,\; \lambda_6 = \frac{1}{ \sqrt{2}} a^\dagger(a + a^\dagger)a \; ,\; \lambda_4 = \frac{1}{ \sqrt{2}} (a^2 + a^{\dagger\,2}) \; ,
\end{equation} 

\ni and

\begin{equation} 
\lambda_3 = \frac{1}{2} (a^2a^{\dagger\,2}- aa^{\dagger\,2} a)\; ,\; \lambda_8 = \frac{1}{ \sqrt{3}} (a a^\dagger  - a^\dagger a) \; . 
\end{equation} 

\ni Note that

\begin{equation} 
[a\,,\, a^\dagger ] =\left(  \begin{array}{ccc} 1 & 0 & 0 \\ 0 & 1 & 0 \\ 0 & 0 & -2  \end{array} \right)
\end{equation} 

\ni is not a canonical commutation relation for bosons, though the canonical annihilation and creation relations (25) hold. Note also the formulae

\begin{equation} 
a = \frac{1}{2} ( \lambda_1 + i \lambda_2) + \frac{1}{\sqrt{2}} ( \lambda_6 + i \lambda_7) \; , \; a^\dagger = \frac{1}{2} ( \lambda_1 - i \lambda_2) + \frac{1}{\sqrt{2}} ( \lambda_6 - i \lambda_7) \; .
\end{equation} 

The (truncated) occupation--number operator (24), appearing naturally in our description of mixing matrix $ U $, tells us that {\it massive neutrinos of three generations, $ \nu_1\, , \,\nu_2\, , \,\nu_3 $, can be characterized by its three eigenvalues} 0,1,2. In fact,

\begin{equation} 
n|n_i \rangle = n_i|n_i \rangle \,,\, a|n_i \rangle = \sqrt{n_i}|n_i - 1 \rangle  \,,\, a^\dagger|n_i \rangle = \sqrt{n_i + 1}|n_i + 1 \rangle \,,\, \langle n_i|n_i \rangle = 1
\end{equation} 

\ni with $ |n_i \rangle = |\nu_i \rangle \;(n_i =0,1,2\,,\, i =1,2,3) $. Here $ \,,\, a|0 \rangle = 0 $ and $ a^\dagger |2 \rangle = 0 $ {\it i.e.}, $ |-1 \rangle = 0 $ and $ |3 \rangle = 0 $, due to the truncation condition (26). We can see from Eqs. (27) that the matrix $ \lambda_5 $, absent from $ U $ in the two--mixing form (2) or (21), involves linearly (in contrast to matrices $ \lambda_2 $ and $ \lambda_7 $) two--step transition operators $ a^2 $ and $ a^{\dagger\, 2}$ mixing directly $ \nu_1 $ and $ \nu_3 $.

Analogical algebraic constructions work also for three generations of other fundamental fermions: charged leptons as well as up and down quarks, but the corresponding parameters $ c_{ij}$ and $ s_{ij}$ take different values. In the representation, where the charged--lepton and up--quark mass matrices are diagonal (and thus the corresponding diagonalizing matrices --- unit), the lepton and quark mixing matrices are, at the same time, the neutrino and down--quark diagonalizing matrices (strictly speaking, in the case of quarks $ V = U^\dagger $ is the conventional mixing matrix). As is well known, in contrast to neutrinos, in the down--quark case no large mixing appears experimentally: the corresponding $ s_{ij}$ are always considerably smaller than $ 1/\sqrt{2} $ (the largest of them is $ s_{12} \sim 0.22)$.

\vfill\eject

~~~~
\vspace{0.5cm}

{\centerline{\bf References}}

\vspace{0.3cm}

{\everypar={\hangindent=0.5truecm}
\parindent=0pt\frenchspacing

{\everypar={\hangindent=0.5truecm}
\parindent=0pt\frenchspacing

1.~Z. Maki, M. Nakagawa and S. Sakata, {\it Progr. Theor. Phys.} {\bf 28}, 870 (1962).

\vspace{0.2cm}

2.~{\it Cf. e.g.} F. Feruglio, {\it Acta Phys. Pol.} {\bf B 31}, 1221 (2000); and references therein.

\vspace{0.2cm}

3.~{\it Cf. e.g.}~J.N.~Bahcall, P.I.~Krastev and A.Y.~Smirnov, {\it Phys. Lett.} {\bf B 477}, 401 (2000); hep--ph/0002293.

\vspace{0.2cm}

4.~Y.~Fukuda {\it et al.} (Super--Kamiokande Collaboration), {\it Phys. Rev. Lett.} {\bf 81}, 1562 (1998) [E. {\bf 81}, 4279 (1998)]; {\bf 82}, 1810 (1999); {\bf 82}, 2430 (1999).

\vspace{0.2cm}

5.~C.~Athanassopoulos {\it et al.} (LSND Collaboration), {\it Phys. Rev. Lett.} {\bf 75}, 2650 (1995); {\it Phys. Rev.} {\bf C 54}, 2685 (1996); {\it Phys. Rev. Lett.} {\bf 77}, 3082 (1996); {\bf 81}, 1774 (1998).

\vspace{0.2cm}

6.~{\it Cf. e.g.} W. Kr\'{o}likowski, {\it Nuovo Cim.} {\bf A 111}, 1257 (1999); {\bf A 112}, 893 (1999), also hep--ph/9904489;  hep--ph/0001023; hep--ph/0004222; and references therein.

\vspace{0.2cm}

7.~W. Kr\'{o}likowski, {\it Acta Phys. Pol.} {\bf B 31}, 663 (2000); and references therein.

\vspace{0.2cm}

8.~M. Appolonio {\it et al.} (Chooz Collaboration), {\it Phys. Lett.} {\bf B 420}, 397 (1998); {\bf B 466}, 415 (1999).

\vfill\eject

\end{document}